\documentstyle[aps,preprint,epsf]{revtex}
\begin{document}
\draft
\title{Finite temperature effects in Coulomb blockade quantum dots and 
signatures  of spectral scrambling}

\author{ Y. Alhassid and S. Malhotra}

\address{ Center for Theoretical Physics, Sloane Physics Laboratory,
     Yale University, New Haven, Connecticut 06520, USA }

\date { August 20, 1999}
\maketitle
\begin{abstract}
  The conductance in Coulomb blockade quantum dots exhibits sharp peaks whose 
spacings  fluctuate with the number of electrons. We derive the 
temperature-dependence  of these fluctuations in the statistical regime and 
compare  with recent experimental results.  The scrambling due to Coulomb 
interactions  of the single-particle spectrum with the addition of an electron 
to  the dot is shown to affect the temperature-dependence of the peak spacing 
fluctuations.  Spectral scrambling also leads to saturation in the temperature 
dependence  of the peak-to-peak correlator, in agreement with recent 
experimental  results. The signatures of scrambling are derived using discrete 
Gaussian processes, which generalize the Gaussian ensembles of random matrices 
to  systems that depend on a discrete parameter -- in this case, the number of electrons in the  dot.

\end{abstract}

\pacs{PACS numbers: 73.40.Gk, 05.45+b, 73.20.Dx, 73.23. -b}

\narrowtext

   A quantum dot is a sub-micron-sized conducting device containing up to 
several  thousand electrons.  In closed dots, the coupling between the dot and 
the  leads is weak and the charge on the dot is quantized 
\cite{Kouwenhoven97}.  The addition of an electron into the dot requires a 
charging  energy of $E_C= e^2/C$ (where $C$ is the capacitance of the dot). 
This  charging energy can be compensated by varying the gate voltage $V_g$, 
leading  to Coulomb blockade oscillations of the conductance versus $V_g$. In 
the  quantum regime (i.e., for temperatures below the mean level spacing 
$\Delta$),  conductance occurs by resonant tunneling, and sharp conductance 
peaks  are observed as a function of $V_g$.

Dots can be fabricated with little disorder such that the electron dynamics 
in  the dot is ballistic. Larger dots are often characterized by irregular 
shape,  resulting in chaotic classical dynamics of the electrons. Such dots 
are  expected to exhibit universal mesoscopic fluctuations which are the 
signature  of quantum chaos.  In particular, the distributions of the 
conductance  peak heights in Coulomb blockade quantum dots at $T \ll \Delta$)) 
were  predicted to be universal, depending only on the underlying space-time 
symmetries  \cite{JSA92}. The measured distributions \cite{Chang96,Marcus96} 
were  found to agree well with theory. The statistics of the peak heights at 
finite  temperatures ($T \sim \Delta$) were also derived recently using random matrix theory (RMT) \cite{Alhassid98}. 
 The measured distributions \cite{Patel98} become narrower 
and  less asymmetric with increasing temperature, in qualitative agreement 
with  theory, although significant deviations were observed at higher 
temperatures,  presumably due to dephasing.

  Another quantity whose statistics was recently studied both experimentally 
and  theoretically is the spacing between successive conductance peaks. In the 
simple  constant interaction (CI) model (where Coulomb interactions are 
included  only as an average charging energy) and for $T \ll \Delta$, a 
(shifted)  Wigner-Dyson peak spacing distribution is expected, but the 
observed  distributions are Gaussian-like 
\cite{Sivan96,Simmel97,Marcus98,Simmel99}.  This has been explained as an 
interaction  effect by numerical diagonalization of a small Anderson model 
with  Coulomb interactions \cite{Sivan96,Berkovits98}. The temperature 
dependence  of the peak spacing statistics was also measured recently 
\cite{Marcus98}.  In this paper we use the finite-temperature theory plus RMT 
to  study the peak spacing fluctuations at temperatures $T \sim \Delta$. We 
find  a rapid decrease of the fluctuations above $T/\Delta \sim 0.5$, in  
agreement  with the experimental results.

Interaction effects beyond the charging energy were not included in the 
finite  temperature theory. They can be treated in a single-particle framework 
within  a mean-field approximation (e.g., Hartree-Fock). Due to charge 
rearrangement,  we expect the spectrum to change or ``scramble'' upon the 
addition  of an electron into the dot \cite{Blanter97,Patel98}. 
 The effects of 
scrambling  on the statistics can be described by a random matrix model that 
depends  on a discrete parameter: the number of electrons on the dot. The 
theory  of discrete Gaussian processes \cite{Attias95} can then be used to 
analyze  the finite temperature statistics of peak spacings and peak 
heights  for various degrees of scrambling. A rescaled parametric distance 
controls  how fast the spectrum is changing, and will be referred to as the 
scrambling  parameter. It was shown that spectral scrambling can lead to 
nearly Gaussian peak spacing distributions at low temperatures \cite{Vallejos98}. In this paper we derive two main signatures of a changing  
spectrum  in the finite temperature statistics: the  less rapid decrease of 
the  spacing fluctuations with temperature for $T/\Delta \agt 0.5$, and the 
saturation  of the number of correlated peaks at higher temperatures 
\cite{MA99}.  The first effect has not been experimentally observed while the 
second  has been qualitatively suggested and observed in Ref. \cite{Patel98}.  
We  also derive a simple expression for the scrambling parameter in terms of 
the  dot's properties.

  At  $T \sim \Delta$, several resonances contribute to each conductance peak 
\cite{Be91} 
\begin{equation}\label{finite-T-g}
G (T,\tilde{E}_F) = \frac{e^2}{h}\, \frac{\pi \bar{\Gamma}}{4 kT} g =  
\sum_\lambda  w_\lambda(T,\tilde{E}_F) g_\lambda\;,
\end{equation}
where $g$ is the dimensionless conductance expressed as a thermal average  
over  individual level conductances $g_\lambda =  2 \bar{\Gamma}^{-1}  
\Gamma_\lambda^l  \Gamma_\lambda^r
   /( \Gamma_\lambda^l + \Gamma_\lambda^r)$. The thermal weight 
$w_\lambda(T,\tilde E_F)$  of a level $\lambda$ (for $T \ll E_C$) is 
given by $w_\lambda 
=  4 f(\Delta F_{\cal N}- \tilde{E}_F)
\langle n_\lambda \rangle_{_{\cal N}}[
1 - f\left(E_\lambda  - \tilde{E}_F \right)]$.
 $\Delta F_{\cal N} \equiv F({\cal N}) - F({\cal N}-1)$ where $F_{\cal N}$ is 
the  canonical free energy of ${\cal N}$ non-interacting electrons on the dot, 
  $\tilde E_F = E_F + e\alpha V_g - ({\cal N}-1/2)E_C$ is an effective Fermi 
energy  ($\alpha$ is the ratio between the plunger gate to dot capacitance and 
the  total capacitance), and $\langle n_\lambda \rangle$ is the canonical 
occupation  of a single-particle level $\lambda$. Both the canonical free 
energy  and occupation are calculated exactly using particle-number projection 
\cite{Alhassid98}.  In the statistical theory, the eigenvalues $E_\lambda$ and 
wavefunctions  $\psi_\lambda$ fluctuate according to the corresponding 
Gaussian  random matrix ensemble. The fluctuations of the partial widths 
$\Gamma_\lambda$  are calculated by relating the widths to the eigenfunctions 
across  the dot-lead interfaces.

\begin{figure}

\vspace{7 mm}

\centerline{\epsffile{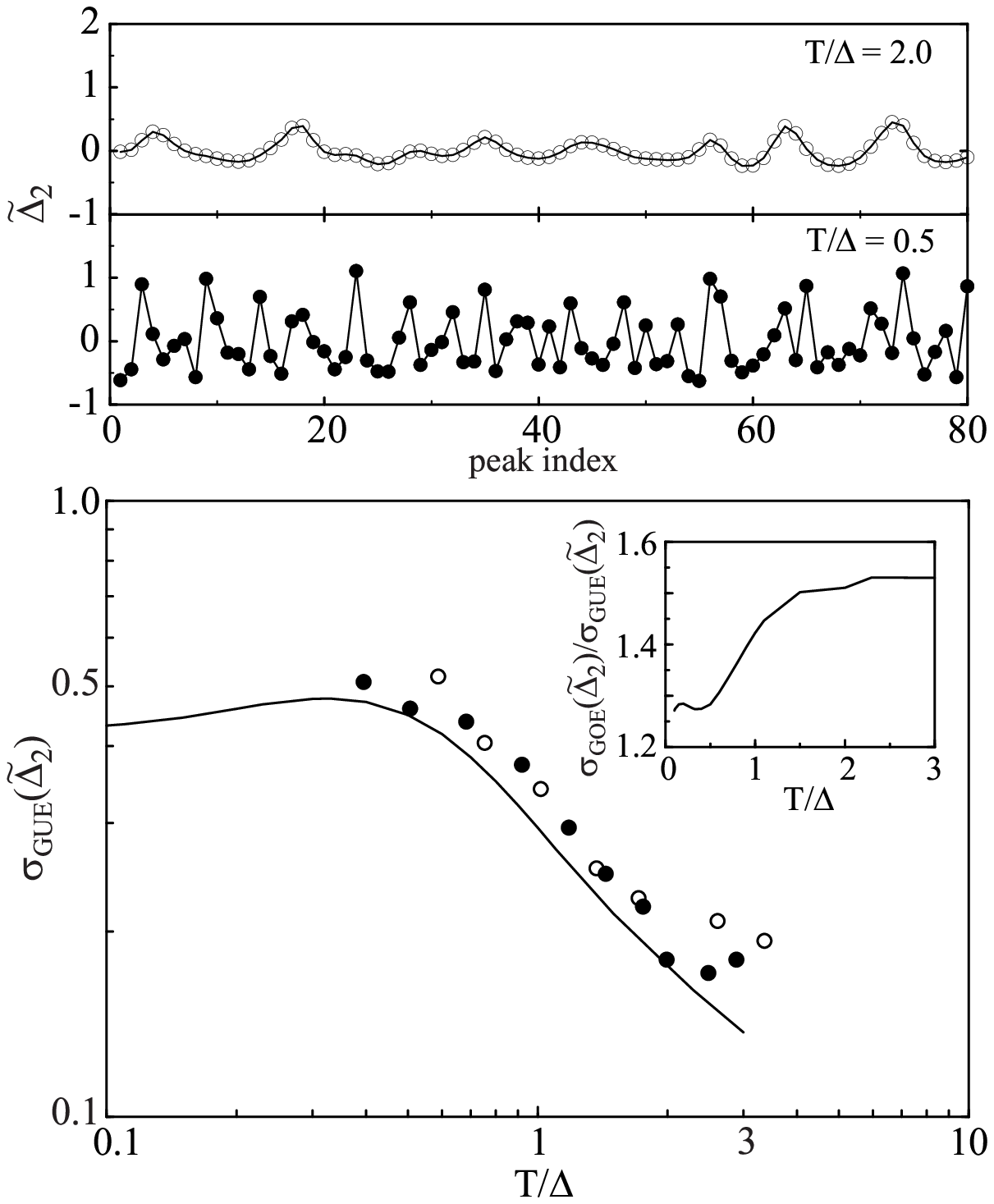}}

\vspace{7 mm}

\caption
{top panel: A typical peak spacing $\tilde \Delta_2 \equiv (\Delta_2 
-\langle\Delta_2\rangle)/\Delta$  sequence obtained from one 
RMT realization of the dot's 
Hamiltonian  at $T/\Delta=0.5$ and $T/\Delta=2$. Notice the reduced 
fluctuations  at the higher temperature. Bottom panel: The standard deviation 
of  the peak spacing $\sigma(\tilde\Delta_2)$ for the GUE statistics as a 
function  of $T/\Delta$ using a log-log scale. The solid line is the 
theoretical  result and the circles are the experimental results of 
\protect\cite{Marcus98}  for two dot configurations. The inset is the 
calculated  ratio $\sigma_{\rm GOE}(\tilde\Delta_2)/\sigma_{\rm 
GUE}(\tilde\Delta_2)$  as a function of $T/\Delta$.
}

\label{fig:sigma-T}

\vspace{7 mm}

\centerline{\epsffile{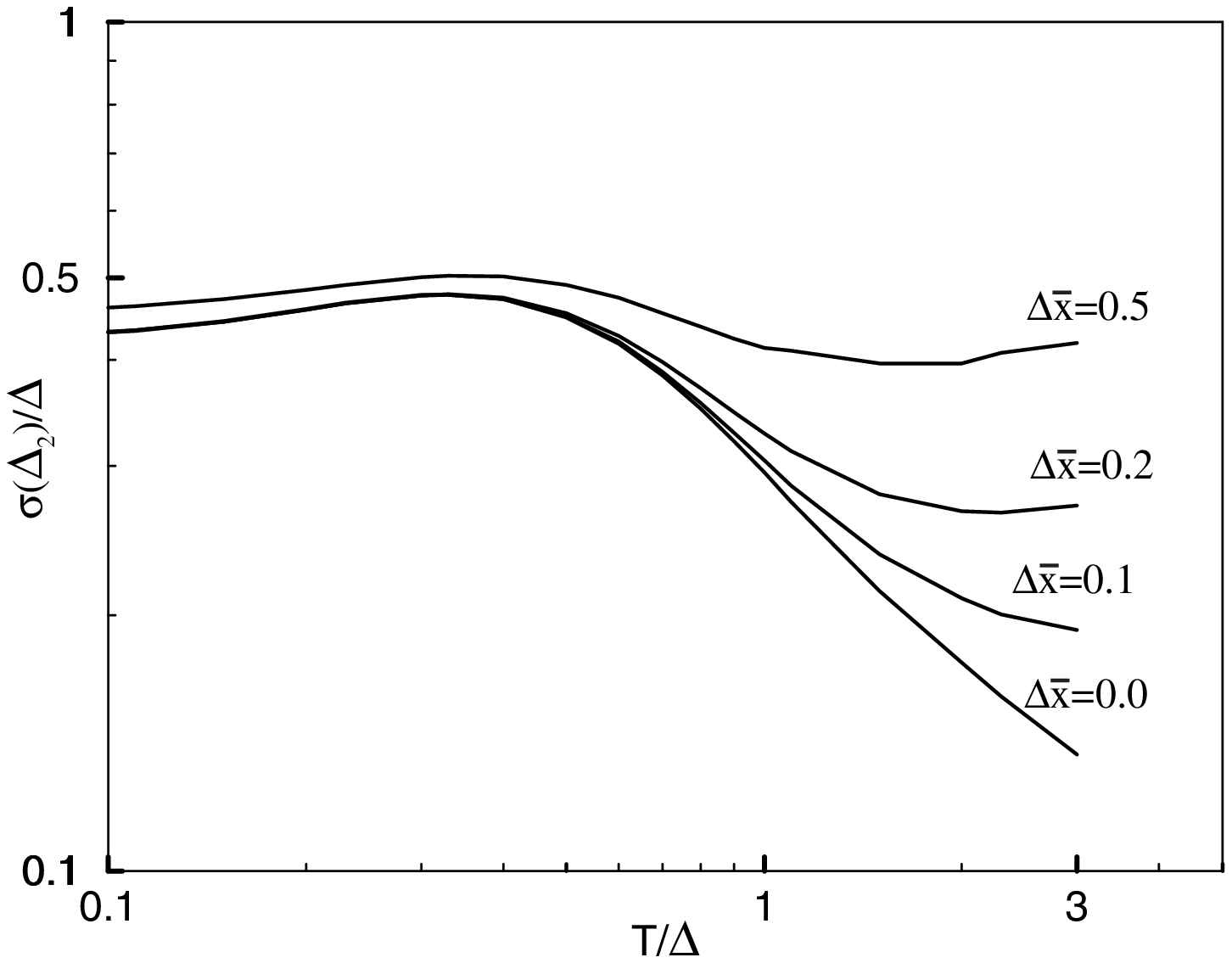}}

\vspace{7 mm}

\caption
{The standard deviation of the GUE peak spacing distribution 
$\sigma(\tilde\Delta_2)$  as a function of $T/\Delta$ (on a log-log scale) for 
different  values of the scrambling parameter $\Delta \bar x$.
}

\label{fig:sigma-x}

\centerline{\epsffile{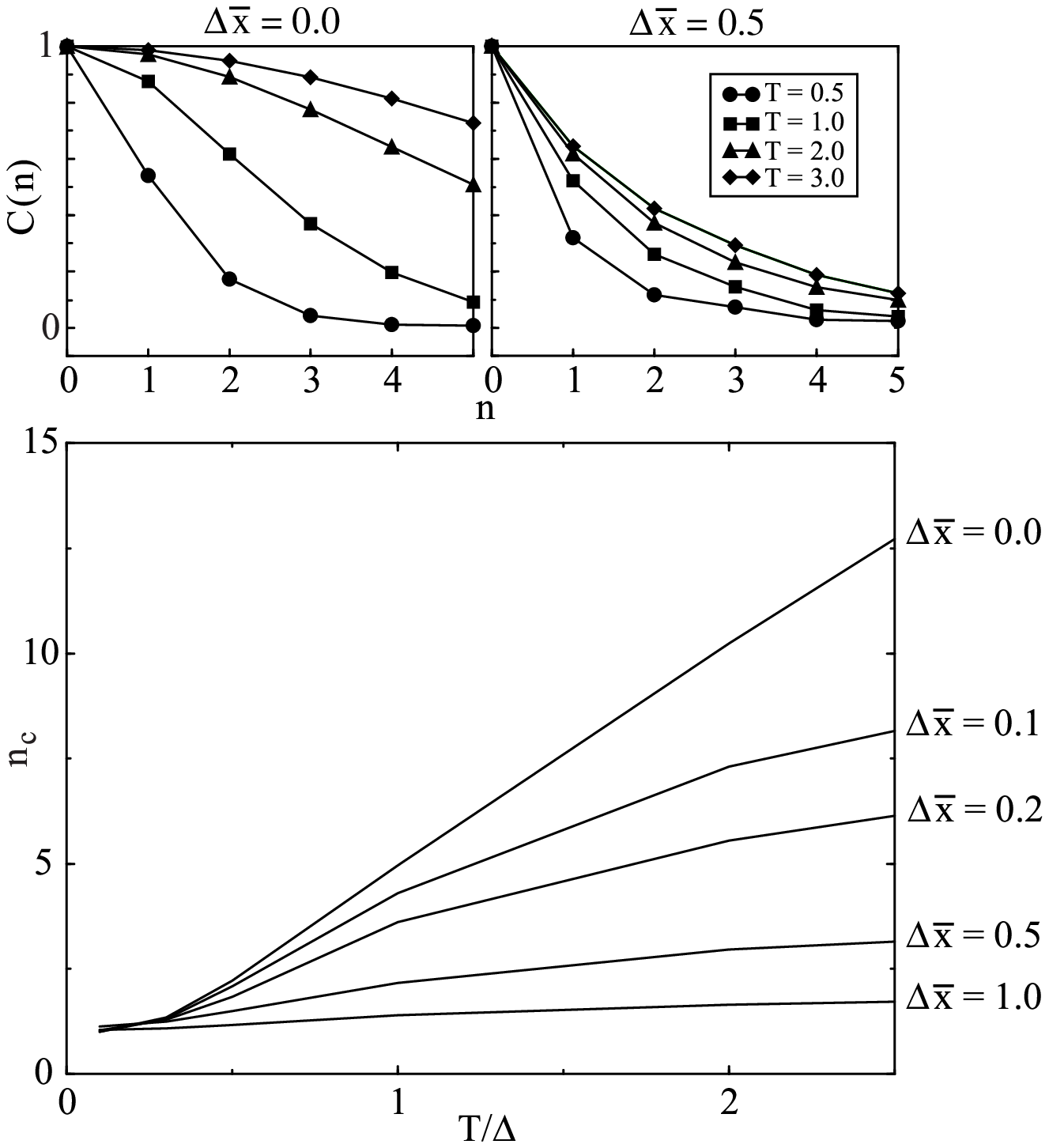}}

\vspace{7 mm}

\caption
{Spectral scrambling and the finite temperature peak-to-peak correlation. Top 
panels:  the peak-to-peak correlator $c(n)$ as a function of distance $n$ 
between  peaks for several temperatures and for a fixed spectrum $\Delta \bar 
x=0$  (top left panel) and a changing spectrum with $\Delta \bar x=0.5$ (top 
right  panel). Bottom panel: the number of correlated peaks $n_c$ (defined as 
the  full width at half maximum of the correlator $c(n)$) as a function of 
$T/\Delta$  for several values of the scrambling parameter $\Delta \bar x= 0, 
0.1,  0.2, 0.5$ and $1$. Notice the saturation of $n_c$ at smaller values for 
larger  values of $\Delta \bar x$ (i.e., smaller dots; see Eq. 
(\protect\ref{Delta-x}).  Compare this figure to the experimental figure 2 of 
Ref.  \protect\cite{Patel98}.
}

\label{fig:peak-corr}

\centerline{\epsffile{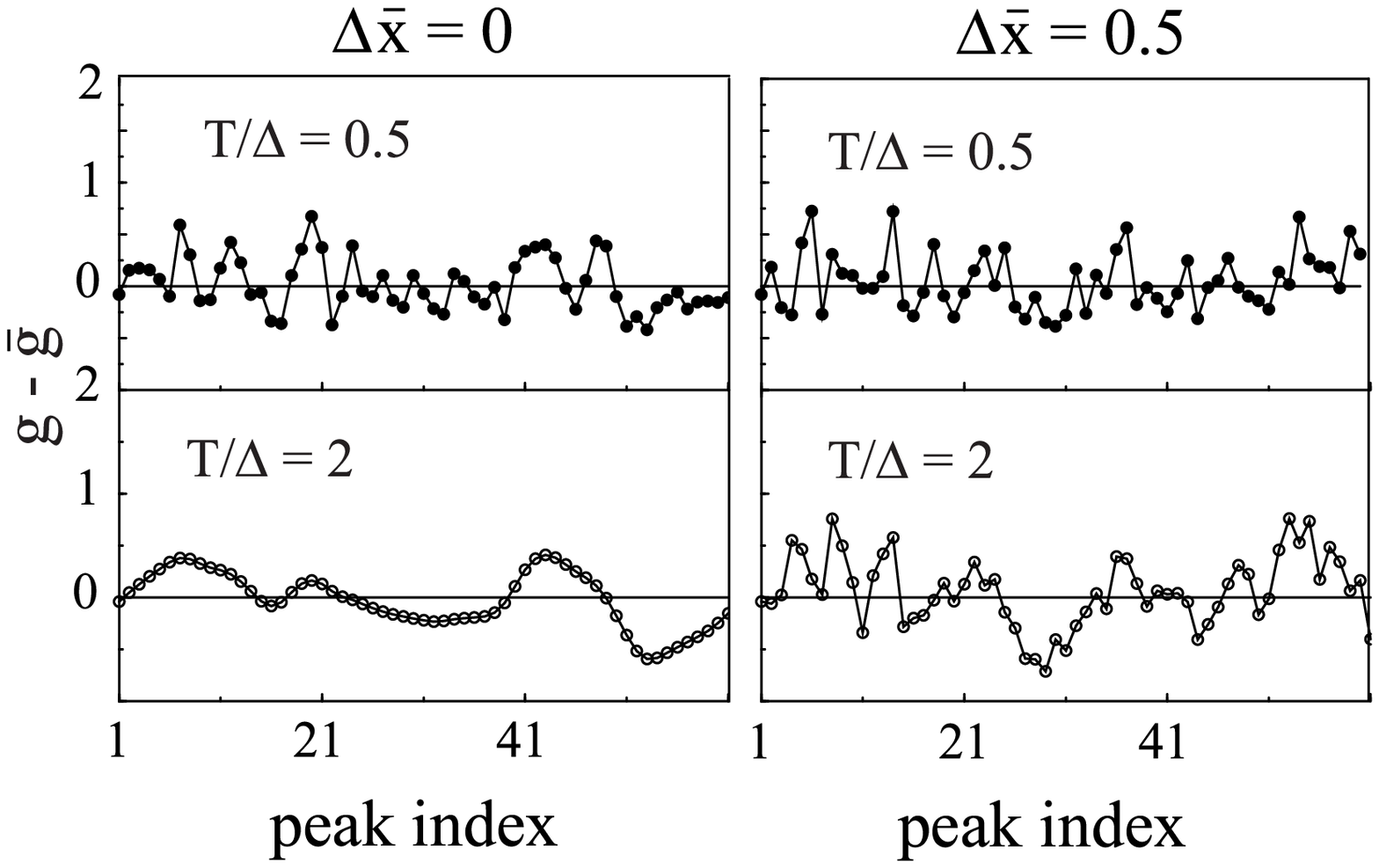}}

\vspace{7 mm}

\caption
{Spectral scrambling and peak height fluctuations. Shown is the peak height  
fluctuation  $g-\bar g$ for a series of peaks generated from one realization 
of  a Gaussian process. Left: $T/\Delta=0.5$ and $T/\Delta=2$ for a fixed 
spectrum  ($\Delta \bar x=0$). Right: $T/\Delta=0.5$ and $T/\Delta=2$ for a 
changing  spectrum ($\Delta \bar x=0.5$)
}

\label{fig:g-fluct}

\end{figure}

 Eq. (\ref{finite-T-g}) was used in Ref. \cite{Alhassid98} to calculate the 
conductance  peak distributions by full RMT simulations, and an approximate 
analytic  expression was derived in the limit where spectral fluctuations are 
ignored.   The finite temperature formulation can also be used to calculate 
the  temperature-dependence of the peak spacing distributions. Unlike the  
peak  height statistics, the peak spacing statistics are sensitive to 
fluctuations  of both the spectrum and the wavefunctions, and full RMT 
simulations  are required. The location of the ${\cal N}$-th peak is 
determined  by finding the value of $\tilde E_F$ for which the conductance  
(\ref{finite-T-g})  is maximal.
Statistics of peak spacings are collected from different successive peaks as 
well  as from different realizations of the dot's Hamiltonian. The peak 
spacings  exhibit less fluctuations at higher temperatures as is demonstrated 
in  the top panel of Fig. \ref{fig:sigma-T}, where the spacings $\Delta_2$ for 
a  typical peak series calculated in one random matrix realization are shown 
at  $T/\Delta =0.5$ and $T/\Delta=2$.   While we do not expect to reproduce 
the  observed functional form (i.e. Gaussian-like) of the peak spacing 
distribution  using a fixed spectrum, it is still meaningful to study the 
temperature-dependence  of the standard deviation of the spacings 
$\sigma(\tilde\Delta_2)$  
(where $\tilde\Delta_2=(\Delta_2-\langle \Delta_2\rangle)/\Delta$). 
The results for the GUE 
statistics  are shown in the bottom panel of Fig.\ref{fig:sigma-T} (solid 
line).  The width shows a slight increase until about $T/\Delta \sim 
0.5$  and then decreases rapidly with increasing $T/\Delta$.  We compare the calculations with recent 
experimental  results \cite{Marcus98} in the presence of a magnetic field 
(circles).  The calculations somewhat underestimate the experimental width but 
describe  well the overall observed temperature dependence. Shown in the inset 
is  the calculated ratio $\sigma_{\rm GOE}(\tilde\Delta_2)/\sigma_{\rm 
GUE}(\tilde\Delta_2)$  which increases as a function of $T/\Delta$. The 
experimental  ratio of $\sim 1.2 - 1.3$ measured at $T \sim 100$ mK is 
consistent  with our theoretical results.

  So far we have taken into account only an average value for the Coulomb 
interaction  (${\cal N}^2 E_C/2$).  Interaction effects can be described 
within  a single-particle framework in the Hartree-Fock (HF) approximation 
\cite{HF99}.  The peak spacing at $T \ll \Delta$ can be expressed as a second 
order  difference of the ground state energy of the dot as a function of 
particle  number \cite{Sivan96}. According to Koopmans' theorem, the change in 
the  HF ground state energy when an electron is added is given by the HF 
single-particle  energy of the added electron
\begin{equation}\label{Koopmans}
{\cal E}_{HF}^{({\cal N}+1)} - {\cal E}_{HF}^{({\cal N})} \approx E_{{\cal 
N}+1}^{({\cal  N}+1)} \;,
\end{equation}
where ${\cal E}_{HF}^{(i)}$ is the HF ground-state energy of the dot with $i$ 
electrons,  and $E^{(i)}_j$ is the energy of $j$-th single-particle state for 
a  dot with $i$ electrons.
The theorem is valid in the limit when the single-particle eigenfunctions are 
independent  of the number of electrons, and is expected to hold for large 
dots  and for interactions that are not too strong. Its validity in numerical 
models  of quantum dots was recently studied \cite{HF99}. The spacing 
$\Delta_2({\cal  N}+1)$ between the ${\cal N}$-th and ${\cal N}+1$ peak is 
then  given by
\begin{equation}\label{Delta}
\Delta_2({\cal N}+1) = E_{{\cal N}+1}^{({\cal N}+1)} - E_{\cal N}^{({\cal 
N})}=   \Delta E^{({\cal N}+1)} + \Delta E_{\cal N} \;.
\end{equation}
$\Delta E^{({\cal N}+1)} \equiv E_{{\cal N}+1}^{({\cal N}+1)} - E_{{\cal 
N}}^{({\cal  N}+1)}$ is the level spacing for a fixed number of electrons 
(${\cal  N}+1$), and $\Delta E_{\cal N} \equiv E_{{\cal N}}^{({\cal N}+1)} - 
E_{\cal  N}^{({\cal N})}$ is the energy variation of the ${\cal N}$-th level 
when  the ${\cal N}+1$ electron is added to the dot.

 The HF Hamiltonian of the dot depends on the number of electrons, and in the 
following  we denote
 by $H(x_{\cal N})$ the Hamiltonian for ${\cal N}$ electrons (in this 
notation   $E_\lambda^{({\cal N})} \equiv E_\lambda(x_{\cal N})$). For a dot 
whose  single-electron dynamics is chaotic we shall assume that $H(x_{\cal 
N})$  is a discrete Gaussian process, i.e., a discrete sequence of correlated 
Gaussian  ensembles of a given symmetry class \cite{Attias95}. Such a process 
can  be embedded in a continuous process  $H(x)$ where the Hamiltonian depends 
on  a continuous parameter $x$ \cite{Wilkinson,AA95}.
A parametric dependence that originates in the dot's deformation as a 
function  of $V_g$ was used to explain the Gaussian-like shape of the peak 
spacing  distribution \cite{Vallejos98}.  Here the parametric dependence is 
assumed  to be mainly due to interaction effects, as recent experimental 
results  indicate \cite{Patel98,Simmel99}.  We further assume that $\Delta x = 
x_{{\cal  N}+1}- x_{\cal N}$ is approximately independent of ${\cal N}$. 
According  to the theory of Gaussian processes, the parametric statistics are 
universal  upon scaling of the parameter by the rms level velocity 
\cite{SA93}.  We denote by $\Delta \bar x$ the variation of the scaled 
parameter  between successive number of electrons. The rms level velocity 
depends  on the symmetry class, and in the following the parameter is always 
scaled  by the GUE rms level velocity.

 A simple Gaussian process is given by \cite{Wilkinson,AA95}
\begin{equation}\label{GP}
H(x) = \cos x H_1 + \sin x H_2 \;,
\end{equation}
where $H_1$ and $H_2$ are uncorrelated Gaussian random matrices. For each 
realization  of the Gaussian process we calculate the single-particle spectrum 
$E_\lambda(x)$  as a function of the parameter. At $T \sim \Delta$, Eq. 
(\ref{Delta})  does not hold. Instead we determine the spacing from the 
location  of successive peaks: the ${\cal N}$-th peak is determined using the 
levels  $E_\lambda(x_{\cal N})$ and wavefunctions $\psi_\lambda(x_{\cal N})$ 
as  explained before, while the ${\cal N}+1$ peak is determined similarly but 
using  a different spectrum $E_\lambda(x_{{\cal N}+1})$ and eigenfunctions 
$\psi_\lambda(x_{{\cal  N}+1})$. Consequently, the spacing $\Delta_2$ depends 
now  on both $T/\Delta$ and $\Delta \bar x$.

 Fig. \ref{fig:sigma-x} shows the standard deviation of the GUE peak spacing 
distribution  $\sigma(\Delta_2)/\Delta$ as a function of $T/\Delta$ (on a 
log-log  scale) for several values of the ``scrambling'' parameter $\Delta 
\bar  x$. As for the $\Delta \bar x=0$ case, we observe a decrease above 
$T/\Delta  \sim 0.5$, except that the decrease is more moderate.  It would be 
interesting  to see whether this signature of spectral scrambling can be 
observed  experimentally.

  Another quantity that is sensitive to spectral scrambling is the 
peak-to-peak  correlator, which is characterized by its full width at half 
maximum  (FWHM), i.e., the number of correlated peaks $n_c$. For a constant 
single-particle  spectrum, $n_c$ is found to increase linearly with 
temperature  since the number of levels contributing to a given peak is $\sim 
T/\Delta$.  However, for a changing spectrum the number of correlated peaks is 
expected  to saturate (as a function of temperature) at a value $\sim m$ that 
measures  the number of electrons needed to scramble the spectrum completely. 
This  effect was observed in the experiment \cite{Patel98}. It can be 
calculated  quantitatively by using the Gaussian process (\ref{GP}) 
\cite{MA99}.  For each value of $\Delta \bar x$ and $T/\Delta$, the 
peak-to-peak  correlations are determined universally. The top left panel of 
Fig.  \ref{fig:peak-corr} shows the calculated peak-to-peak correlator $c(n)$ 
at  several temperatures for a constant spectrum (i.e., $\Delta \bar x=0$), 
where   the correlator width is seen to increase with temperature. The right 
inset  shows the correlator $c(n)$ for the same temperatures but for a 
changing  spectrum $\Delta \bar x= 0.5$, where its width is seen to saturate 
at  higher temperatures. The bottom panel of Fig. \ref{fig:peak-corr} shows 
the  number of correlated peaks $n_c$ versus $T/\Delta$ for several values of 
$\Delta  \bar x$.  $n_c$ saturates at a smaller value of $m$ as $\Delta \bar 
x$  gets larger, i.e., when the single-particle spectrum scrambles faster. To 
further  illustrate this effect,  we show in Fig. \ref{fig:g-fluct} the peak 
height  fluctuation $g-\bar g$ for a series of peaks using one realization of 
the  GP (\ref{GP}) at $T/\Delta=0.5$ and $2$. For a fixed spectrum ($\Delta 
\bar  x=0$) we observe a significant increase of the correlations at the 
higher  temperature (left panels), while for a changing spectrum ($\Delta \bar 
x=0.5$)  the correlations do not change much with temperature (right panels).

  Experimentally it was found that, in a smaller dot, $n_c$ saturates at a 
smaller  value $m$ \cite{Patel98}. This suggests  faster scrambling (i.e., 
larger  $\Delta \bar x$) in the smaller dot. We can derive an expression for 
the  scrambling parameter $\Delta \bar x$ in terms of the dot's properties by 
relating  the  parametric approach to the microscopic HF-RPA approach of 
\cite{Blanter97}.   In the latter the fluctuations in the variation $\Delta 
E_{\cal  N}$ of the ${\cal N}$-th level upon the addition of an electron (see 
Eq.  (\ref{Delta})) are dominated by charge that is pushed to the surface, and 
are  estimated to be $\sigma^2 (\Delta E_{\cal N}) \sim \beta^{-1} 
\Delta^2/g$,  where $g$ is the dimensionless Thouless conductance. In the 
parametric  approach we find, using perturbation theory \cite{AA95},
$\sigma^2 (\Delta E_{\cal N})= 2 \beta^{-1} \Delta^2 (\Delta \bar x)^2$. 
Comparing  the two expression for $\sigma^2(\Delta E_{\cal N})$ we obtain
\begin{equation}\label{Delta-x}
\Delta \bar x \propto g^{-1/2} \sim {\cal N}^{-1/4}
\;.
\end{equation}
The symmetry class parameter $\beta$ drops out in this relation. Indeed we 
expect  $\Delta \bar x$ to depend only on the dot's properties irrespective of 
the  presence or absence of a magnetic field. Relation (\ref{Delta-x}) is 
valid  in the regime where $\sigma^2(\Delta E_{\cal N})$ is linear in $(\Delta 
\bar  x)^2$, i.e., for $\Delta \bar x \alt 0.3$ (see Fig. 2 of Ref. 
\cite{Attias95}).  The second estimate in (\ref{Delta-x}) is for a ballistic 
dot where $g \sim {\cal N}^{1/2}$.  Relation (\ref{Delta-x}) confirms that $\Delta \bar x$ is larger for the 
dot  with a smaller number of electrons.   Notice the qualitative similarity 
between  the theoretical results of Fig. \ref{fig:peak-corr} and the 
experimental  results in Fig. 2 of Ref. \cite{Patel98}.

In conclusion, we used the statistical theory of Coulomb blockade quantum 
dots  to calculate the temperature-dependence of the peak spacing 
fluctuations.  Statistical scrambling of the spectrum upon adding an electron 
to  the dot affects the temperature-dependence of both the peak spacing 
fluctuations  and the peak-to-peak correlator.

This work was supported in part by the Department of Energy grant
No. DE-FG-0291-ER-40608.

\end{document}